\documentclass[preprint2]{aastex}
\usepackage{epsfig,graphicx}
\begin{document}

\title{Two New Ejecta-Dominated Galactic Supernova Remnants: 
G337.2$-$0.7 and G309.2$-$0.6}
\author{
Cara E. Rakowski\altaffilmark{1,}\altaffilmark{2}, 
John P. Hughes\altaffilmark{1,}\altaffilmark{2}, \and
Patrick Slane\altaffilmark{3}
}
\altaffiltext{1}{Department of Physics and Astronomy, Rutgers The State
 University of New Jersey, 136 Frelinghuysen Road, Piscataway NJ 08854-8019;
 E-mail: rakowski@physics.rutgers.edu and jph@physics.rutgers.edu
}
\altaffiltext{2}{
Also Service d'Astrophysique, L'Orme des Merisiers, 
CEA-Saclay, 91191 Gif-sur-Yvette Cedex France
}
\altaffiltext{3}{Harvard-Smithsonian Center for Astrophysics, 60 Garden Street,
Cambridge MA 02138  % ; slane@head-cfa.harvard.edu
}

\begin{abstract}

We present the analysis of new X-ray observations of two Galactic
radio supernova remnants (SNRs), G337.2$-$0.7 and G309.2$-$0.6. Both
remnants exhibit line-rich thermal spectra that require highly
non-solar elemental abundance ratios. In each case, Si and S are
unequivocally overabundant compared to solar based purely on the
measured equivalent width of the K$\alpha$ line complex.  Detailed
nonequilibrium ionization spectral analysis of these remnants, using a
single temperature, single ionization age model, confirm the overabundances,
yielding Si, S and Ar abundances many times their solar values: 3 to 5
times solar for G337.2$-$0.7 and greater than 3 to 10 times solar for
G309.2$-$0.6. We also find that for both remnants, Ne and Mg are
underabundant relative to Si, and for G309.2$-$0.6, Ca and Fe as well
are relatively underabundant.  This is the first detection of
metal-rich, non-solar abundance material in these remnants,
identifying them as young ejecta-dominated SNRs.  Further support for
their youth comes from the distances that we estimate from the fitted
column densities (less than 15~kpc for G337.2$-$0.7, and $4 \pm 2$~kpc
for G309.2$-$0.6). From the implied sizes of the remnants at these
distances we infer relatively young ages (2000 to 4500 years for
G337.2$-$0.7, and 700 to 4000 years for G309.2$-$0.6). Finally we note
that no compact object has yet been found within G337.2$-$0.7, but an
X-ray point source is evident within the radio shell of G309.2$-$0.6,
although its nature remains unknown.

\end{abstract}


\section{Introduction}

Supernovae (SNe) are the birthplace of almost all metals that exist in
the universe. Models of the explosive nucleosynthesis that occurs in
SNe are used to study the evolution of the interstellar medium,
external galaxies, and even clusters of galaxies. Young supernova
remnants (SNRs), whose emission is still dominated by the ejecta, are
the best test for our models of nucleosynthesis.  However, these
models have primarily been tested in the ensemble, against the
abundances of large regions, such as the solar neighborhood, where
those abundances are assumed to be the product of many representative
supernovae.  Direct measurements of the nucleosynthetic yields of
individual supernova, using the abundances found in their ejecta, has
been much more limited.  Cas A, which has been extensively studied
across the wavebands (e.g., Chevalier \& Kirshner~1979; Fesen, Becker,
\& Blair~1987; Douvion, Lagage, \& Cesarsky~2000; Hughes et~al.~2000)
is more the exception than the rule.

Considering that the Galactic SNe rate is widely believed to be a few
per century, there should have been on the order of 5--10 Galactic SNe
since the last historical one observed by Kepler in 1604. Additionally
there must be a sizeable population of SNRs less than $\sim$2000 years
old or so. However, prior to {\it ASCA} there were only a handful of
young remnants known and they fell into two classes: the remnants of
young historical SNe (e.g. Tycho, Kepler) and oxygen-rich SNRs
(e.g. Cas A, G292.0$+$1.8) classified based on their optical
characteristics. Studies with the {\it Advanced Satellite for
Cosmology and Astrophysics} ({\it ASCA}) have added a third class of
presumably young remnants: those for which optical evidence of ejecta
is absent, but whose X-ray spectra show enhanced, non-solar
abundances, and hence are likely to be ejecta-dominated (e.g. W49B in
the Galaxy, Fujimoto et~al. 1995; or N103B in the LMC, Hughes
et~al.~1995).  Clearly, identifying and studying new X-ray
ejecta-dominated SNRs will be vital for understanding the SNR
population and the process of nucleosynthesis.

G337.2$-$0.7 and G309.2$-$0.6 are part of our ongoing project
surveying small Galactic supernova remnants in an attempt to identify
more ejecta-dominated SNRs through their X-ray emission.  G337.2$-$0.7
and G309.2$-$0.6 were first identified as SNR candidates based on
their possible non-thermal radio emission (Green 1974). These
identifications were confirmed in the radio in the MOST supernova
remnant catalogue (Whiteoak \& Green 1996). G309.2$-$0.6 was the
subject of a detailed radio study by Gaensler, Green, \& Manchester
(1998) (hereafter GGM) using the Australia Telescope Compact Array
(ATCA).  The radio images of G337.2$-$0.7 and G309.2$-$0.6 are given
in Figures~1a and~2a.  G337.2$-$0.7 shows a barely resolved, possibly
shell-like morphology, with a diameter of $\sim$6~arcminutes and an
overall radio surface brightness at 843 MHz of $11\times
10^{-21}$~W~m$^{-2}$~Hz$^{-1}$~sr$^{-1}$ (Whiteoak \& Green 1996).
G309.2$-$0.6 exhibits a more complex morphology in the radio with
evidence for 2 shells, diameters 12 and 15~arcminutes, and a possible
jet-like feature to the northeast (GGM) which appears as a faint wisp
in our reproduction. The SNR has a surface brightness at 843 MHz of
$5.4\times 10^{-21}$~W~m$^{-2}$~Hz$^{-1}$~sr$^{-1}$ (Whiteoak \& Green
1996). GGM find a distance for G309.2$-$0.6 of between 5.4$\pm$1.6 and
14.1$\pm$0.7 kpc using H{\small I} absorption measurements.  GGM's
lower limit was found based on absorption towards the SNR, and the
upper limit was constrained by the lack of a positive velocity
absorption feature which was present in other nearby sources,
indicating that the remnant must be closer than those sources.

These SNRs are positionally coincident with bright sources from the
{\it ROSAT} All-Sky Survey (RASS) (Voges et~al.~1999).  G337.2$-$0.7
corresponds to source 1RXS~J163931.4$-$475019, with a count rate of
$0.05 \pm 0.02$ counts s$^{-1}$, and G309.2$-$0.6 is close to source
1RXS~J134651.1$-$624843 with a count rate of $0.06 \pm 0.02$ counts
s$^{-1}$.  These measurements indicated the presence of soft X-ray
emission from these remnants; our {\it ASCA} observations were
required to determine the nature of this emission.

The {\it ASCA} X-ray observations of these remnants provide new
insight into their physical properties. The {\it ASCA} band (0.7$-$10
keV) spans emission lines of Ne, Mg, Si, S, Ar, Ca, and Fe, allowing
for examination of the abundances, temperature, density and ionization
state in the X-ray emitting region, as well as the line-of-sight
absorbing column density, although individual emission lines are, in
general, not resolved by {\it ASCA}. This first examination of the X-ray
spectra from these remnants reveals line-rich thermal emission with
strong indications of highly non-solar elemental abundance ratios.

In section 2 we outline the data reduction, background subtraction and
general image analysis. In section 3 we present the spectroscopy and
nonequilibrium ionization (NEI) analysis, along with the resulting
abundances. Section 4 discusses the distances, ages, and
nucleosynthesis models. Finally, we summarize our results and highlight
areas for further investigation in section 5.

\section{Data Reduction and Image Analysis}

The primary X-ray observations used in this work were taken with
instruments aboard {\it ASCA}, which has two Solid State Imaging
Spectrometers (SIS~0 and SIS~1) which each consist of four CCD chips,
and two Gas Imaging Spectrometers (GIS~2 and GIS~3) (For details
concerning the {\it ASCA} satellite see Tanaka, Inoue \& Holt,
1994). Our observations were taken in 1997 March for G337.2$-$0.7 and
1999 April for G309.2$-$0.6, both late in the life of {\it ASCA},
after significant cosmic-ray induced degradation of the CCDs had
occurred. Two CCD mode was chosen to minimize the effect of residual
dark current due to radiation damage (the amount of dark current
increases with the readout time which scales with the number of
chips), while still covering the full extent of each remnant.

The GIS and SIS data were screened with the standard {\it ASCA}
criteria.\footnote{Interested readers are referred
to the {\it ASCA} Guest Observer Facility, and the {\it ASCA} Data
Reduction Guide at http://heasarc.gsfc.nasa.gov/docs/asca/abc/abc.html
for more details.}  Both light-leak and contamination from cosmic rays
are a concern.  For the GIS data, rise-time information was used to
reject non X-ray events.  To determine if the standard screening were
sufficient for our SIS data, we checked to see if there were any
significant differences in the spectra for medium versus high
bit-rate, night versus day, different cutoff rigidities, bright versus
faint mode, and between the two SIS instruments for each source
studied.  Since we found no significant secular changes in the
spectra, we concluded that indeed the standard screening was
reasonable. After screening, the exposure times for G337.2$-$0.7 were
$\sim$14.6~ksec in each of the SIS instruments (bright and faint modes
combined, both medium and high bit-rate).  Likewise the good exposure
times for G309.2$-$0.6 were $\sim$15.4~ksec with SIS~0, but only
$\sim$4.6~ksec with SIS~1.

Both the GIS and SIS data were used in the image analysis. Due to its
higher temporal resolution ($\sim$0.016~s) only the high-bit-rate GIS
data were used for timing analysis.  For the spectral analysis we
chose to focus on the higher resolution SIS data.  The GIS spectra are
qualitatively similar, but detailed spectral analysis of these data is
more difficult because, at the time of this work, significant
uncertainties exist in the GIS gain (as much as 45~eV shifts near the
Si K$\alpha$ line complex; Ken Ebisawa, personal communication). For
an emission-line dominated spectrum, such as the ones here, this
uncertainty would strongly influence our results, and hence we choose
to use only the SIS spectra.

\subsection{G337.2$-$0.7, image analysis and background subtraction}

Images of G337.2$-$0.7 in various wavebands are shown in Figure 1.  In
Figure 1a, we compare the MOST radio image with contours from the
overall (0.7-10.0 keV) {\it ASCA} SIS X-ray band.  We also present the
{\it ASCA} SIS X-ray image separated into ``soft'' (Figure 1c) and
``hard'' (Figure 1d) bands, where the soft band (0.7$-$1.5 keV) covers
L-shell lines of Fe and Ca, and K-shell lines of Ne and Mg, and the
hard band (1.5$-$10.8 keV) includes emission from the K-shell lines of
Si, S, Ar, Ca and Fe. Both bands include continuum emission. These
images were made by combining the SIS~0 and SIS~1 images in those
energy bands, correcting for the exposure time in any given pixel and
smoothing the image by convolving it with a gaussian of $\sigma \sim
0.8^{\prime}$.  In the X-ray band G337.2$-$0.7 appears as a
featureless source $\sim$$6^{\prime}$ in diameter, comparable to the
MOST radio image, and therefore nearly unresolved by {\it ASCA}. There
is no significant change in the size of the remnant with energy,
especially at high energies, hence there is no evidence for the
presence of a hard, pulsar-powered, synchrotron nebula within the
remnant. We do not show the GIS images since they do not reveal any
new information beyond that shown by the SIS.

The soft band image (Figure 1c) shows another source toward the
northeast of the SNR, which we designate AX~J16398$-$4746.  Figure 1b
overlays the soft band contours on the Digitized Sky Survey (DSS)
image from the UK Schmidt Telescope plates. There is a bright star
(HD~149901, spectral class F5, $m_{V}=7.4$) that is positionally
coincident with the soft X-ray source.  This source has an X-ray count
rate of $(5.0 \pm 0.6)\times 10^{-3}$~counts~s$^{-1}$ in a circular
region of radius 1.7$^{\prime}$, yielding a greater than 6$\sigma$
detection of the source.  The X-ray flux is $f_{x} \sim 5\times
10^{-13}$ergs~cm$^{-2}$~s$^{-1}$ (over the 0.3 to 3.5 keV band). The
X-ray to optical flux ratio of $\log (f_{x}/f_{v}) \sim 4$ is
consistent with that of an F5 star according to
Maccacaro~et~al.~(1988) making plausible the identification of
HD~149901 as the X-ray source.

The spectrum for G337.2$-$0.7 was extracted from a circular region
with radius 3.7$^{\prime}$ which excluded the soft emission from the
HD star, but was sufficiently large to account for the point spread
function (PSF) of the {\it ASCA} SIS and the extent of the source.
The spectra from the two SIS instruments were combined and then
grouped into bins containing at least 25 counts each so that the
$\chi^{2}$ statistic could be used in fitting the spectrum. We
followed the standard {\it ASCA} prescription for generating a
response function for combined SIS data.  Routine background
subtraction of the average cosmic and detector background was also
done using publically available high Galactic latitude fields.  For
the SNR we find a count rate of $0.099 \pm 0.002$~SIS~counts~s$^{-1}$,
where only the high Galactic latitude background (with rate
0.018~SIS~counts~s$^{-1}$) has been subtracted.

For G337.2$-$0.7, additional background X-ray emission from the
Galactic ridge was found to be present.  To model this emission, a
source-free control region of the same size as the SNR region was
extracted from elsewhere within the SIS field of view.  Using an
absorbed powerlaw model, which was a good fit to the data, with
best-fit column density $N_{\mathrm H}\sim 3.9\times
10^{21}$~atoms~cm$^{-2}$ and spectral index, $\alpha_{p} \sim 1.3$, we
find a surface brightness of $3\times
10^{-7}$~ergs~s$^{-1}$~cm$^{-2}$~sr$^{-1}$ in the 0.5-10~keV band.
This is consistent with the surface brightness of Galactic ridge
emission found near G328.4$+$0.2 by Hughes, Slane, \&~Plucinsky
(2000), and the Galactic ridge emission found with the {\it GINGA}
satellite by Yamauchi (1991). This absorbed powerlaw model for our
control region was then included as an additional component in the
spectral fits for the SNR, in order to account for the Galactic ridge
emission contained in the SNR spectrum. Our model implies a count rate
of $0.016$~SIS~counts~s$^{-1}$ from the Galactic ridge in the SNR
region.

\subsection{G309.2$-$0.6}

In Figure 2 we show images of G309.2$-$0.6 in various wavebands.
Figure 2a shows the {\it ATCA} 1.3 GHz image with broad-band X-ray
contours of the {\it ASCA} GIS observations that were exposure
corrected, summed and smoothed with a $\sigma = 1^{\prime}$ gaussian. The
field of view of the GIS extends slightly beyond that shown in
Figure~2a; however no significant X-ray emission was detected beyond
that shown. In particular the region of the radio jet toward the north
is not detected.

For the {\it ASCA} SIS X-ray images of G309.2$-$0.6, we separated the
emission into the band in which extended emission was significant
(1.0-3.2~keV, Figure~2d), and energies at which an unresolved source
in the field of view, 1WGA J1346.5$-$6255, dominates the emission
(0.7-1.0~keV and 3.2-10.0~keV, Figure~2c). These images were exposure
corrected, combined and smoothed with a  $\sigma = 0.8^{\prime}$
gaussian. Both the SNR (0.14~SIS~counts~s$^{-1}$, see
below) and the unresolved source ($0.027 \pm 0.001$~SIS~counts~s$^{-1}$ in
a $\sim 1^{\prime}$ region) are detected at much
greater than the 6$\sigma$ level.  As we will show in section 3.2, it
appears that effectively all of the emission in the 0.7-1.0~keV and
3.2-10.0~keV bands comes from the unresolved source.

In Figure~2b, we show the {\it ROSAT} PSPC X-ray contours overlayed on
the DSS greyscale image of this region.  For all
instruments the position of the unresolved source is consistent with
the optical cluster NGC~5281 as shown in Figure~2b and previously
noted for the {\it ROSAT} data by GGM. Modeling the spectrum of the
point source with an absorbed powerlaw, column density $N_{\mathrm H}
= 2.6^{+1.8}_{-1.4}\times 10^{21}$~atoms~cm$^{-2}$, and spectral index
$\alpha_{p}=1.5 \pm 0.3$, we find that the flux from this source in
the 0.3$-$3.5~keV band is $\sim 1.3 \times
10^{-12}$~ergs~cm$^{-2}$~s$^{-1}$. This flux is consistent with that
expected from HD~119682, the third brightest star in NGC~5281, an O9,
$m_{V}=7.89$ star, using the relation of Maccacaro et~al.~(1988),
although it is on the X-ray bright side of the distribution for such a
stellar type. In Figure~2b this star is the brightest one within the
highest {\it ROSAT} contour.  It is also possible that the point
source is a compact object orbiting one of the stars in the cluster
NGC~5281.

There is even a more intriguing possibility, that the point source is
related to the SNR itself. Given that the absorbing column density
found is similar to that towards the remnant (see section~3.2
and Table~2), it is reasonable to expect the point source to be at the
same distance as the remnant. The photon index of the point source is broadly
consistent with a pulsar or synchrotron nebula interpretation.  To
search for pulsations from this source we chose a region $1^{\prime}$
in radius in the GIS data to maximize the signal to noise, yielding
177 counts total in the two GIS instruments, with as much as 20\%
contamination from the SNR and cosmic x-ray background. No significant
peak was found in the Fourier transform of the GIS lightcurve in the
2$\times 10^{-3}$~Hz to 32~Hz frequency range. The lightcurve was
epoch folded over periods near each of the largest peaks in the
Fourier transform, but no pulsations were found above even the
1~$\sigma$ confidence limit. However, due to the small number of
counts, this only limits the pulse fraction to be less than $\sim$85\%
over this frequency range at the 99.99\% confidence level (see Leahy
et~al.~1983).

Due to the broad PSF of {\it ASCA} the emission from
the unresolved source spreads over the field of view and significantly
contaminates the remnant emission. As a consequence, the spectral
analysis of G309.2$-$0.6 requires special careful treatment.

For G309.2$-$0.6, spectra from two source regions were extracted,
one circular region of radius 0.95$^{\prime}$ which contained the most
obvious emission from the point source, and another which excluded
this region, and contained the full extent of the remnant as seen in
the 1.0-3.2~keV band, with a radius of 5.6$^{\prime}$.  The spectra
and response functions from the two SIS instruments were combined in
the same manner as for G337.2$-$0.7, and routine background
subtraction was done using the high Galactic latitude background
files.

For G309.2$-$0.6, we also extracted control regions to model the
Galactic ridge background, but we found only one region with
significant counts after subtracting the cosmic background and this
region's emission could easily be attributed to instrumental
scattering from the supernova remnant and the point source. Therefore
no additional Galactic ridge background was subtracted for
G309.2$-$0.6.

The primary source of contamination to the {\it ASCA} SIS spectrum of
G309.2$-$0.6 comes from the point source. To model how the
PSF of {\it ASCA} behaves at this particular
position on the chip as a function of energy, we chose to investigate
archival data of an isolated point source observed at a similar
position on the chip in the same observational configuration as our
data.  The cataclysmic variable source, SS Cyg ({\it ASCA}
seq. no. 30001000), was found to be a reasonable match, most notably
in the respect that the source is the same distance from the chip
gap. This source is also bright, so there are sufficient counts to 
use it to model the PSF.

Using regions nominally identical in size and position to those of the
G309.2$-$0.6 point source and remnant regions, we found the ratio of
counts in the inner smaller region to the outer region for each energy
bin.  We then used this ratio to model the contamination from the
G309.2$-$0.6 point source into the SNR region. The contamination
spectrum was modeled with an absorbed powerlaw and this powerlaw was
used as one of the components for the model of the SNR emission. It
should be noted that the ratio was not strongly dependent on photon
energy, and that merely renormalizing the original point source
powerlaw model by an appropriate amount gave a reasonable fit for the
modeled SNR contamination spectrum.

Various sources of systematic error in the model of the point source
contamination were investigated, that could either increase ($+$) or
reduce ($-$) the contamination from the point source into the SNR
region. These were as follows: (1) small changes in the positions of
the extraction regions for the point source ($\pm$6\%) and the SNR
($+$6\%) based on the SS Cyg data, (2) contamination of the point
source by SNR emission that falls in the point source region
($-$2\% to $-$10\%)), and (3) including a Galactic ridge background component
($+$14\%).  To account for these possible sources of error, we have
investigated both the nominal model for the contamination discussed
above as well as a model in which the contaminating flux has been
decreased by 10\%. However, we can safely ignore a model with the
contaminating flux increased by even as little as a few percent since
this would overpredict the emission from the remnant region below
1~keV and above 3.2~keV.

\section{Spectral Analysis}

We modeled the emission from each remnant with a nonequilibrium
ionization (NEI) model. The NEI model used in this analysis is the
single-temperature, single-timescale model from Hughes \& Singh (1994)
which assumes that the SNR emission can be described by an average
temperature $kT$ and ionization timescale $n_{e}t$, which
parameterizes the departure from ionization equilibrium.  The elements
included in this study are H, He, C, N, O, Ne, Mg, Si, S, Ar, Ca, and
Fe whose baseline solar abundances are as given by Raymond \& Smith
(1977): 12.0, 10.93, 8.52, 7.96, 8.82, 7.92, 7.42, 7.52, 7.2, 6.9,
6.3, 7.6 dex respectively. We note that the inferred abundances are
model dependent, however the basic conclusions should remain
unchanged.

We explored models with three different sets of elemental abundances in
our NEI spectral fits. First, the abundances of all the elements were
fixed at their solar values to determine a baseline model fit.  Next,
the overall metallicity was allowed to vary freely, i.e. the H and He
abundances were fixed at their solar values, and the abundances of all
higher Z elements were allowed to vary, but with abundances fixed
relative to one another.  In our final model, which we refer to as the
varying abundances model, C, N, and O were fixed to their solar
abundances, Ne and Mg were varied together, while Si, S, Ar, Ca and Fe
were each individually allowed to vary freely. The rationale behind
this prescription is the following: the emission lines of C, N, and O
are below or at the lower limits of the detected energy range so that
they have little or no effect on the results.  The emission lines from
Ne and Mg, although detectable, are highly absorbed so they are weak
and independent determination of each abundance separately is not well
constrained. For G337.2$-$0.7 the observed ratio of Ne to Mg was
similar to the solar ratio when both were allowed to be free, but
neither was tightly constrained in this method, so we tied their
abundances together. For G309.2$-$0.6 the abundance of Ne was not
individually constrained in any way, so it could have been set to
solar abundance. However, since the preferred ratio of Ne to Mg
abundance was close to solar, for G309.2$-$0.6, we chose to link the
Ne abundance to the Mg abundance as was done for G337.2$-$0.7.  Si, S,
Ar, Ca, and Fe abundances were allowed to vary freely because they
have detectable emission lines in the {\it ASCA} SIS X-ray band, even
for a highly absorbed source. 

\subsection{G337.2$-$0.7}
Figure~\ref{fig:spectrum} shows the combined SIS spectra from
G337.2$-$0.7 with the best-fit spectral model (see below). The most
obvious features of this spectrum are the emission lines of Si, S and
Ar, as indicated on figure~\ref{fig:spectrum}.  The lines themselves
are each an amalgam of multiple emission lines of different ionic
species, whose individual lines are not resolved at this spectral
resolution. Due to degradation of the CCD He-like and H-like lines are
no longer resolvable from each other.  For instance, the feature
around 1.9~keV contains K$\alpha$ ($ {\mathrm n}=2 \rightarrow
{\mathrm n}=1$) lines predominantly of He-like and H-like Si.  This
line blend is quite strong.  Using a simple bremsstrahlung continuum
and three gaussian lines with an absorbing column we find that the
equivalent widths (EWs) for the Si, S and Ar K$\alpha$ line complexes
of G337.2$-$0.7 are about 660, 570, and 320 eV, respectively ($\pm
50$~eV at 1$\sigma$).  In comparison, the maximum equivalent widths
for a solar abundance plasma, EW$_{\rm max, \odot}$, for Si, S, and Ar
for temperatures between 0.4 and 5.0 keV, any ionization timescale, and
any absorbing column density are 550, 350, 400 eV respectively. From
this we can immediately see that Si and S are significantly enhanced
in G337.2$-$0.7.  Additionally, the temperatures, timescales and
column densities at which the EW$_{\rm max, \odot}$ are found are not
consistent with the data. For example the modeled energies of the
lines are different from those in the gaussian line fits indicating
that the modeled ionization state is not correct and the line ratios
between the different atomic species are dissimilar from those found
in G337.2$-$0.7.

For G337.2$-$0.7, the various spectral models yield the following.
With all the abundances fixed to their solar values the best-fit
$\chi^{2}$ is 170.5 (73 degrees of freedom), clearly not a good fit.
The model with a single variable metallicity prefers an extremely
large metallicity (7 times solar) and does slightly better at fitting
the spectra than the solar abundance model with a $\chi^{2}$ of 155.7
(72 degrees of freedom).  Both these models have $kT\sim$ 0.7 keV,
$n_{e}t\sim 5 \times 10^{4}$ cm$^{-3}$ yr, and $N_{\mathrm H}\sim 4
\times 10^{22}$ atoms cm$^{-2}$. With similar temperature, timescale
and absorbing column density (see Table~\ref{ta:best-fit2}), the
varying abundances model provides a much improved description of the
spectrum. It yields a best-fit $\chi^{2}$ of 103.1 (67 degrees of
freedom). This reduction in $\chi^{2}$ implies a F$_{\chi}$-stat
confidence level for the addition of six new parameters (compared to
the solar model) of 99.5\%.  Si, S and Ar are all significantly
overabundant compared to their solar values confirming what we found
above from the line equivalent widths, whereas the best-fit abundance
of Ne and Mg is zero.

Using the varying abundances model, we mapped the allowed region of
parameter space for temperature and ionization timescale.  As given in
Table~\ref{ta:best-fit2}, the 1$\sigma$ range of $kT$ and $n_{e}t$ is
$kT = 0.82-0.89$~keV and $n_{e}t = (2.1 -
8.0)\times10^{4}$~cm$^{-3}$~yr. Note that the upper limit on $n_{e}t$
is simply the value at which the equilibrium ionization fractions are
attained. We see that a low temperature and near equilibrium
ionization fractions are preferred.  The H density, corresponding to
the best-fit emission integral is $1.4 \pm 0.3D^{-1/2}_{10
\rm{kpc}}$~atoms~cm$^{-3}$ scaled to a nominal distance of 10~kpc (see
section~4). The absorbing column density is high, $(3.5 \pm
0.3)\times10^{22}$~atoms~cm$^{-2}$.

In figure~\ref{fig:kTvary} we present a concise summary of the
spectral fit results, including the variation with temperature of the
best-fit $\chi^{2}$, the ionization timescale, the best-fit abundance
of Si (relative to solar), and the abundances of the other elements
relative to Si. The range plotted corresponds to the 3$\sigma$ allowed
range on $kT$.  In figure~\ref{fig:abund} we present the best fit values
and 1$\sigma$ limits on the relative metal abundances in the form,
(Z/Si)$\times$[(Si/H)$_{\odot}$/(Z/H)$_{\odot}]$.

According to figures~\ref{fig:kTvary}~and~\ref{fig:abund}, Si, S, Ar,
and Fe are all constrained at the 1$\sigma$ level to be greater than
1.7 times their solar abundances, and Si is constrained to be at least
3.2 times its solar abundance. Furthermore, the overabundance of Si,
S, Ar and Ca hold true throughout the entire allowed 3$\sigma$ range
on $kT$. In addition to these enhanced abundances, further evidence
for the unusual nature of the emission comes from the non-solar ratio
of Ne and Mg to Si.  Although the absolute abundances of Ne and Mg are
allowed to be greater than solar, the ratio of their abundance to Si
is constrained to be less than 0.2 times the solar ratio at
1$\sigma$. Unfortunately the situation is not so clear for the Fe
abundance, since, in addition to the best-fit abundance of Fe (which
is super-solar), there is a region of parameter space for which its
abundance is zero.

\subsection{G309.2$-$0.6}

Figure~\ref{fig:spectrum309} shows the combined SIS spectra of
G309.2$-$0.6 with the best-fit spectral model. This spectrum exhibits
very strong emission lines from Si, S and Ar. As before, using a
simple bremsstrahlung continuum and three gaussian lines with an
absorbing column we find EWs for Si, S, and Ar of $1160\pm 100$ eV,
$1280\pm 100$ eV, and $510\pm 200$ eV, respectively, for the nominal
contamination model (to repeat, EW$_{\rm max, \odot}$: 550, 350, and
400 eV for Si, S, and Ar).  Furthermore, even if contamination from
the point source is ignored entirely, the EWs of Si and S are $640\pm
50$~eV and $550\pm50$~eV, still greater than the EW$_{\rm max, \odot}$
values.

G309.2$-$0.6 also requires non-solar relative abundances.  Models with
solar abundances or a single variable metallicity do not provide a
good fit to the spectrum, for either point-source contamination
model (i.e. nominal or one with 10\% less contamination).  In the case
with nominal contamination, the best-fit solar abundance model and
single variable metallicity model have $\chi^{2}$ (d.o.f.)  values of
111.0 (59) and 100.4 (58) respectively with $kT\sim$ 0.5~keV,
$n_{e}t\sim 6 \times 10^{3}$ cm$^{-3}$ yr, and $N_{\mathrm H}\sim 5
\times 10^{22}$ atoms cm$^{-2}$. This should be compared to a
$\chi^{2}$ of 49.4 (53) for the varying abundance model (see
Table~\ref{ta:best-fit4}). This large reduction in $\chi^{2}$ corresponds
to an F$_{\chi}$-stat confidence level for the addition of these six
new parameters (compared to the solar abundance model) of 99.5\%.

Our spectral fits for G309.2$-$0.6 require extremely enhanced
abundances. The best-fit value for the single variable metallicity is
effectively infinite ($>$1000): in other words, the minimum $\chi^{2}$
is found when emission from H and He becomes negligible. Similarly,
high values are also found for the varying abundances model (see
Table~\ref{ta:best-fit4}). At a minimum, Ne, Mg, Si, S and Ar are
highly over-abundant, with lower limits of 2.0, 2.0, 10.9, 11.9 and
2.9 respectively (where this lower limit derives from the 1$\sigma$
errors on the model with 10\% less contamination than the nominal
case).  In fact a pure-metal plasma, composed of only those elements
with obvious emission lines in this band is easily allowed within the
1$\sigma$ limits of either contamination model.  The relative
elemental abundances quoted in Table~\ref{ta:best-fit4} are also
highly non-solar. In particular, the abundances of Ne, Mg, Ca and Fe
relative to Si are all less than half the solar ratios.

For G309.2$-$0.6, where a pure metal plasma of just Ne, Mg, Si, S, Ar,
Ca and Fe is allowed, it is apparent that the relevant abundance
parameters to study are the abundance ratios to Si, not to H.  In
Figure \ref{fig:kTvary309}, we plot $\chi^{2}$, the ionization
timescale, and the abundance ratios with Si as a function of $kT$ over
a wide range, for both contamination models. We note that throughout
this range in $kT$ the Si abundance is greater than 34 and 11.9 at the
1$\sigma$ level for the nominal and 10\% less contamination models
respectively and moreover, must be greater than 6.5 and 3.9 respectively
at the 99\% confidence level. The pattern of variations here is
similar to that for G337.2$-$0.7: Fe is only allowed for low
temperatures, where Ne and Mg are at a local minimum, and the other
relative abundances remain fairly constant throughout the range of
parameter space. Si, S, Ar, Ne and Mg are all highly overabundant
compared to solar over the entire range of $kT$.  In
Figure~\ref{fig:abund309} we present the best fit values and 1$\sigma$
limits on the relative metal abundances in the form,
(Z/Si)$\times$[(Si/H)$_{\odot}$/(Z/H)$_{\odot}]$, for both the nominal
and the 10\%-less contamination model. This figure also illustrates
the dominance of Si and S emission and relative underabundance of Ne,
Mg, Ar, Ca, and Fe.

The effect of the point source contamination on the derived abundances
is important to note.  As one might expect, since less of the
continuum emission has been accounted for, the 10\%-less contamination
model yields lower abundances relative to H than the nominal
contamination model does. However the abundances relative to Si are
somewhat higher. Yet, the pattern of the relative abundances remains
qualitatively the same, and the enhanced abundances persist even
without any contamination model. We note that the 10\%-less
contamination model does a marginally better job of fitting the data
($\chi^{2}$ of 48.6 vs. 49.4) but the difference in $\chi^{2}$ is not
significant.

Verification of our model for the point-source contamination of
G309.2$-$0.6 comes from examining the {\it ROSAT} data on this
remnant. The PSF of {\it ROSAT} is much finer than {\it ASCA} and
hence we expect very little contamination of the SNR emission from the
point source.  After background subtraction, the {\it ROSAT} spectrum
for the remnant shows very little emission below 1 keV (count rate =
0.01$\pm$0.004 counts s$^{-1}$, between 0.5$-$1.0 keV compared to
0.07$\pm$0.006 counts s$^{-1}$ for 0.5$-$2.0 keV), which agrees with
our model that point source contamination is causing effectively all
the emission below 1~keV for the {\it ASCA} data.  Modeling the {\it
ROSAT} SNR emission with the best-fit model for just the SNR emission
in {\it ASCA}, gives a slightly better $\chi^{2}$ than a model
assuming that what we define as contamination from the point source is
actually SNR emission. Unfortunately, the $\chi^{2}$ difference is not
great enough to allow us to discriminate between these two
cases. However we can conclude that our point source contamination
models are fully consistent with the {\it ROSAT} data.

The dynamical parameters for G309.2$-$0.6 are less well constrained
than for G337.2-0.7, because of the complication of the
contaminating point source.  Nevertheless the best-fit values of $kT$,
$n_{e}t$, and $N_{\mathrm H}$ are all quite reasonable. G309.2$-$0.6
appears to have a somewhat higher temperature than G337.2-0.7 (1-2 keV
vs. 0.8-0.9 keV). The ionization timescales for the two SNRs are
consistent with each other and the 1$\sigma$ ranges allow for
equilibrium ionization for both remnants.  G309.2$-$0.6 has a low
absorbing column density, N$_{\rm H}$,
$(0.3-1.0)\times10^{22}$~atoms~cm$^{-2}$, similar to that found for
the unresolved source, indicating that they are at approximately
the same distance. The number density of hydrogen atoms, corresponding
to the best-fit emission integral and a nominal distance of 5.4~kpc is
very low, $n_{\mathrm{H}}$=$0.02^{+0.03}_{-0.01} D^{-1/2}_{5.4
{\mathrm kpc}}$ ~atoms~cm$^{-3}$, reflecting the fact that hydrogen is
not the dominant element in the X-ray emitting plasma as it would be
under usual solar abundance conditions.

\section{Discussion}

Given the absorbing hydrogen column density measured from our X-ray
spectra, it is possible to estimate the distances to G309.2$-$0.6
and G337.2$-$0.7. Following Chen et~al. (1999), given a hydrogen
column density, $N_{H}$, the relation between column density and
optical color excess, $N_{H} = 5.9 \times 10^{21} \langle E_{B-V}
\rangle $ (Predehl \& Schmitt 1995) and the extinction per unit
distance, $\langle E_{B-V} \rangle /d$, in the direction of a remnant
one can estimate the distance to that remnant. For the extinction per
unit distance, we use the contour diagrams of Lucke (1978). Since
these measurements were for stars within 2~kpc of the sun, and the
density of absorbing material is undoubtedly greater near the Galactic
center, any distance estimates greater than 2~kpc in the direction
towards the Galactic center should be viewed as upper limits.

For G309.2$-$0.6, with $N_{\mathrm{H}} = (0.7 \pm 0.3)\times10^{22}$,
$\langle E_{B-V} \rangle /d \sim 0.3$~mag~kpc$^{-1}$, we find $d \leq
4 \pm 2$~kpc. The distance thus found for G309.2$-$0.6, is consistent
with the lower range of distances found by GGM, 5.4$\pm$1.6~kpc, which
was found from \ion{H}{1} absorption toward the remnant itself. This
distance is consistent with that of the \ion{H}{2} region, RCW80 lying
$\sim 20^{\prime}$ north of G309.2$-$0.6, as well as the location of
the Scutum-Crux spiral arm of the Galaxy. Furthermore, for a distance
of 5.4~kpc, GGM estimate an age of less than 4000 years and size of
$\sim$9.5~pc in radius which is more consistent with its
ejecta-dominated nature than a larger distance would be.

For the point source in G309.2$-$0.6, 1WGA J1346.5$-$6255, which has
$N_{\mathrm H} = 2.6^{+1.8}_{-1.4}\times 10^{21}$~atoms~cm$^{-2}$ we
find $d = 1.5 \pm 1$~kpc, broadly consistent with the X-ray
estimated distance to the SNR, but also consistent with the distance
to the cluster NGC 5281 (1.3~kpc, Moffat \& Vogt 1973).

The $N_{\mathrm{H}}=(3.5 \pm 0.3) \times 10^{22}$ atoms cm$^{-2}$ and
$\langle E_{B-V} \rangle /d \sim 0.4$~mag~kpc$^{-1}$ of G337.2$-$0.7
yield a distance upper limit of 15.5$\pm 1.5$~kpc. This is much
further than the applicability of the extinction per unit distance
estimate, but is useful as an upper limit. In the direction of
G337.2$-$0.7 the measured total column density is rather high,
$\sim$$5\times 10^{22}$~atoms~cm$^{-2}$; much of this material is
likely to be distributed in molecular clouds and other dense regions
in the Perseus, Scutum-Crux and Norma spiral arms and the 5 kpc
molecular ring. Since a single large molecular cloud along the line of
sight could be sufficient to account for all the X-ray absorption we
see (e.g., Corbel~et~al.~1999), we can, at this time, only surmise
that G337.2$-$0.7 is somewhere between the near and far sides of the
Norma spiral arm at $\sim$5~kpc and $\sim$15~kpc. Measurements of the
intensities and velocities of CO emission towards this remnant would
allow us to determine the distribution of material along the
line-of-sight.  Combined with the X-ray column density, we could then
estimate the remnant's distance, as has been recently done by
Corbel~et~al.~(1999) for SNR G337.0$-$0.1.  In addition, radio
measurements of \ion{H}{1} absorption to G337.2$-$0.7 should be
pursued as another constraint on its distance. For purposes of
estimation in this work, we use a nominal distance of 10~kpc.

Given our distance estimates, we can place rough limits on the ages of
the remnants. By assuming free-expansion at some typical velocity
(5000~km~s$^{-1}$) we can place a lower limit on the time since the
explosion. For G337.2$-$0.7, distances of 10 and 15~kpc yield a lower
limit of between 1700 and 2600 years. For G309.2$-$0.6 a distance of
4$\pm$2~kpc corresponds to a lower limit on the age of between 700 and
2000 years. Since our abundances indicate that these remnants are not
yet in the Sedov phase, a nominal Sedov solution provides upper limits
on the ages (see for instance Hamilton, Sarazin, \& Chevalier 1983;
Hughes, Hayashi, \& Koyama 1998). This is because a Sedov solution
would imply that the outer shock has been decelerated by the ambient
medium more than we would expect for a remnant still dominated by
emission from ejecta.  For both remnants we assume an ambient density,
$n_{0} = 0.2$~cm$^{-3}$, and explosion energy, $E_{0} = 10^{51}$~ergs;
derived ages scale as $D^{5/2}\,(n_0/E_0)^{1/2}$. At the upper limits
of our distances we find that G337.2$-$0.7 is less than $4500$ years
old and G309.2$-$0.6 is less than $4000$ years old (the latter as
previously reported by GGM). These age estimates are nicely consistent
with the ejecta-dominated nature of these remnants, and furthermore
any distances closer than these rough upper limits would imply
correspondingly younger ages. However, since they are based on merely
nominal values of the velocity, explosion energy and density, they
should not be over-interpreted.  Future observations with {\it
Chandra} or {\it XMM-Newton} will be able to determine the blastwave
temperature and emission measure (as opposed to the ejecta temperature
and emission measure found here) so that we can solve for the actual
$E_{0}$, $n_{0}$ and shock velocity, and directly constrain the ages
of these remnants.

One would like to use the abundance ratios found in our spectral fits
to determine the types of SN that formed each of these
remnants. However, SNRs are spatially complex and undergo significant
temporal evolution so that we do not expect all the ejecta to be at
the same thermodynamic state, the reverse shock to have thermalized
all the ejecta at the same time, or even for the ejecta to be
uniformly distributed. In fact recent experience with {\it Chandra}
data from Cassiopeia A (Hughes et~al. 2000) has shown that for this
Type II SNe, different nucleosynthetic burning products from distinct
layers of the progenitor star are seen with differing thermodynamic
states and at widely separate positions in the remnant. Therefore our
fitted abundances, based on a single temperature, single timescale
model, at best give only a rough indication of the total
nucleosynthetic yields.

With the preceding caveats in mind, in Figures~\ref{fig:abund}
\&~\ref{fig:abund309}, we compare the best-fit relative abundances for
G337.2$-$0.7 and G309.2$-$0.6, to nucleosynthesis models of Type Ia
and II explosions from Nomoto et~al. (1997), Thielemann, Nomoto, \&
Hashimoto (1996) and Woosley \& Weaver (1995). Only the few cases that
gave reasonable fits are shown.  The abundance ratios for
G337.2$-$0.7, are a decent match to the SN Ia model W7 (Nomoto
et~al. 1997). Type Ia produce very small quantities of low Z elements
such as Ne and Mg, and more Si-group elements such as S, Ar and Ca,
which fits well with the G337.2$-$0.7 abundance patterns.  The Fe
abundance is less than that expected from the total nucleosynthetic
yield, but this could simply be because the entire Fe-rich core has
not yet been shocked. A low-mass SN II core-collapse model, for
instance the 13~$M_{\odot}$ model of Thielemann, Nomoto, \& Hashimoto
(1996) (labeled TNH~13~$M_{\odot}$), is the next best fit for
G337.2$-$0.7. The shortcoming of this model is its overprediction of
the Ne and Mg abundances. However, since the low abundance of Ne and
Mg is more strongly constrained than the Fe abundance, we slightly
prefer the SN Ia model.

For G309.2$-$0.6, the case is less clear. The closest model is a
15~$M_{\odot}$ SN II model of Woosley \& Weaver (1995) (labeled
WW~15~$M_{\odot}$), chosen for its low Fe abundance and high S
abundance. The presence of significant Ne and Mg also argues for a SN
II model.  However, it is clear that a global nucleosynthetic yield is
an unreasonable model since there is no evidence for Fe.  Determining
the type of the progenitor of this remnant will require further
investigation.  If for instance the point source in the X-ray image of
G309.2$-$0.6 were confirmed to be associated with the remnant and
shown to be a compact remnant that would clinch its origin as a
core-collapse SN.

\section{Conclusions}

The enhanced abundances derived from the strong X-ray emission lines
of highly ionized Si, S and Ar reveal the ejecta-dominated nature of
both G337.2$-$0.7 and G309.2$-$0.6. The EWs of the Si and S lines for
both remnants are greater than the maximum possible EWs for solar
abundance plasmas with temperatures between 0.4 and 5.0 keV.  In
addition the prominence of these lines compared to the Ne and Mg
lines, requires non-solar abundance ratios.

Using nonequilibrium ionization thermal plasma spectral models we
derive abundances for these remnants that are highly
non-solar. Over the entire range in temperatures and ionization
timescales, the abundances of Si, S, Ar and Ca are many times the
solar values for G337.2$-$0.7 and for G309.2$-$0.6 a pure metal plasma
of only Ne, Mg, Si, S and Ar (i.e., no H or He) is allowed. In
addition, the elemental abundances are not well described by a single
multiplicative factor times their solar abundances. For G337.2$-$0.7,
Ne, Mg, and possibly Fe are significantly underabundant relative to
the Si-group elements.  For G309.2$-$0.6 Ne, Mg, Ar, Ca and Fe are all
underabundant compared to Si and S.  We thus conclude that
G337.2$-$0.7 and G309.2$-$0.6 are new examples of ejecta-dominated
remnants.

Furthermore our spectral fits provide us with a measure of the column
densities to each remnant, which when coupled to estimates of the
absorption as a function of distance along a given line of sight,
provides distance estimates for both remnants. For G309.2$-$0.6 we
find a distance of $4 \pm 2$~kpc consistent with both GGM's estimate
from \ion{H}{1} absorption lines, and the position of the Scutum-Crux
spiral arm. For G337.2$-$0.7 we find an upper limit on the distance of
15~kpc, placing it no further than far side of the Norma spiral
arm. Using these distances, we can place lower limits on the ages of
the remnants by assuming free-expansion. G309.2$-$0.6 is at least 700
to 2000 years old, while G337.2$-$0.7 is greater than 1700 to 2600
years old. A nominal Sedov solution, $E_{0} = 10^{51}$~ergs, $n_{0} =
0.2$~cm$^{-3}$ yields upper limits on the ages of 4500 years for
G337.2$-$0.7, and 4000 years for G309.2$-$0.6.  These age estimates
corroborate our finding that both remnants are young and
ejecta-dominated.

We also draw attention to the X-ray point source 1WGA~J1346.5$-$6255,
apparent within the radio shell of G309.2$-$0.6. We found no pulsations
for this source in the 0.002~Hz to 32~Hz frequency range, but this
only limits the pulsed fraction to be less than $\sim$85\%. Although
we were able to estimate its distance using the absorbing column
density, we can neither exclude the possibility that it is
associated with the SNR nor with the foreground open cluster
NGC~5281. Higher spatial and temporal resolution observations as are
possible with the {\it Chandra X-ray Observatory} are vital to resolve
the nature of this source as well as to unambiguously distinguish
between the emission from this point source and the SNR itself.

Continued investigation of ejecta-dominated SNRs such as G337.2$-$0.7
and G309.2$-$0.6 is crucial.  Higher spectral resolution observations
as will be possible with {\it XMM-Newton} and {\it Chandra} will allow
further constraints on both the abundances and the thermodynamic state
of the shocked plasma in these remnants. Further, the breadth of
possible investigations that take advantage of the 0.5$^{\prime
\prime}$ spatial resolution of {\it Chandra} is staggering. Now it is
possible to study the abundances of individual knots of ejecta and
compare them with models of explosive nucleosynthesis from different
layers in the exploding SN and thereby trace the dynamics of the
explosion, as we have already begun for Cassiopeia A (Hughes
et~al.~2000). We can also isolate the spectrum of small-scale features
in a remnant (e.g., the outermost blast-wave) and thereby study how
SNe shocks impart energy to the ISM, heat electrons and ions, generate
cosmic rays, and so on (see Hughes, Rakowski, \& Decourchelle 2000).
Thus studies of young ejecta-dominated remnants, like those presented
here, with the new generation of X-ray observatories, have the
potential to make a significant impact on a number of important
astrophysical questions.

\begin{acknowledgements}
We thank Bryan Gaensler for use of the ATCA data on G309.2$-$0.6, as
well as helpful commentary and Paul Plucinsky for useful discussions
on the scientific content of this work. CER and JPH would like to
thank Monique Arnaud for support and hospitality during the course of
this work.  This research has made use of the NASA Astrophysics Data
System, the CDS SIMBAD database and archival X-ray data provided by
the HEASARC at NASA Goddard Space Flight Center. Also shown were
Digitized Sky Survey images, which were found using {\it SkyView} also
provided by the HEASARC. Partial support was provided by NASA grant
NAG5-6420.

\end{acknowledgements}

\clearpage

\figcaption[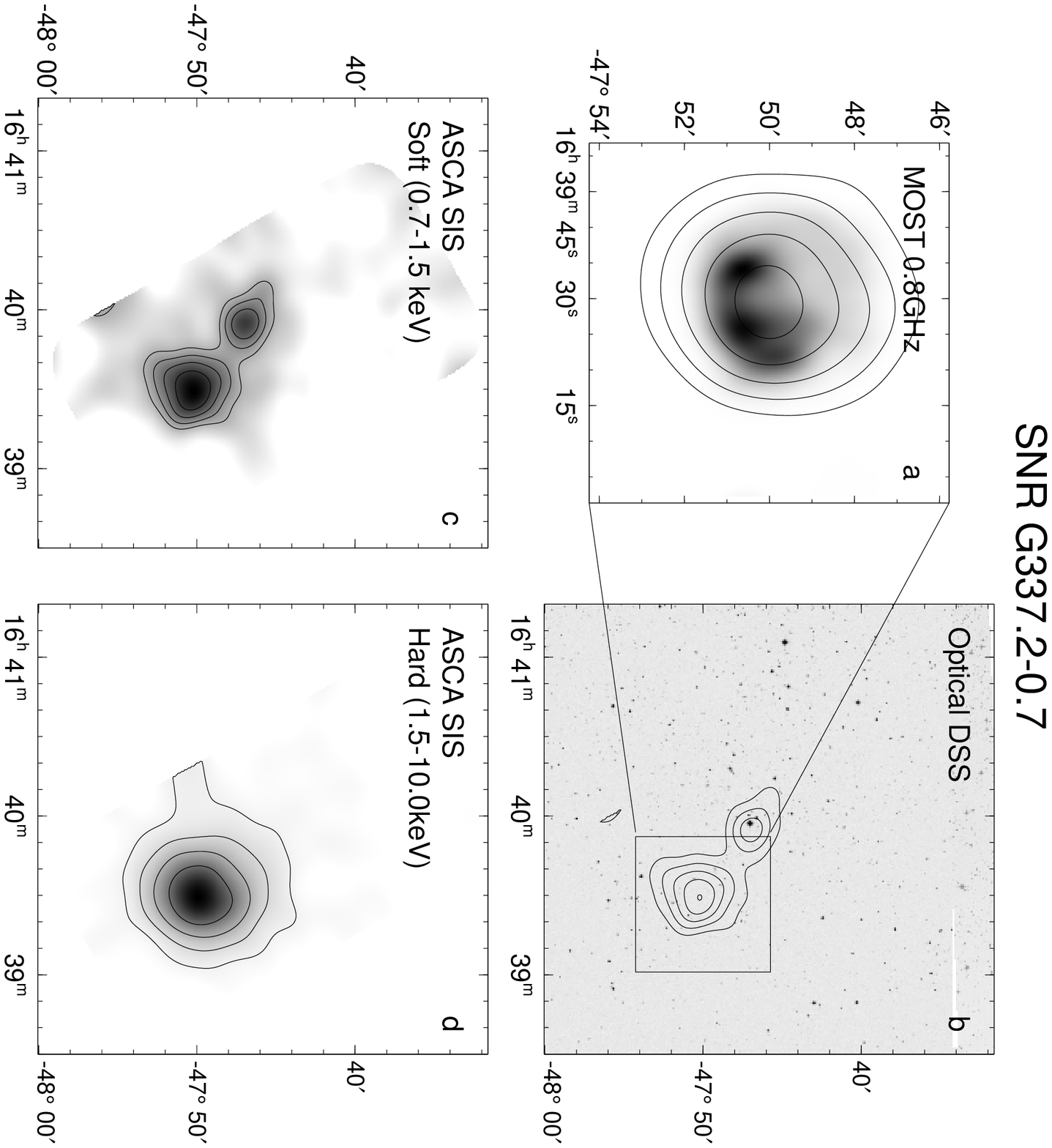]{Image of G337.2$-$0.7 in various bands. All
grey-scales are displayed linearly from the background level to the
maximum, all contours are linearly spaced from 3$\sigma$ above the
background to 90\% of the maximum. Part a: MOST 0.8~GHz image in
grey-scale, overlayed with 0.7-10.0~keV SIS contours; levels (in
10$^{-3}$ counts s$^{-1}$ arcminute$^{-2}$) 1.61, 2.24, 2.86, 3.48,
4.10. Part b: Digitized Sky Survey image overlayed with 0.7-1.5~keV
SIS contours. The brightest star within the X-ray contours is
HD~149901. Part c: SIS 0.7-1.5~keV; levels (in 10$^{-4}$ counts
s$^{-1}$ arcminute$^{-2}$) 3.61, 4.46, 5.31,6.16, 7.01. Part d: SIS
1.5-10.0~keV; levels (in 10$^{-4}$ counts s$^{-1}$ arcminute$^{-2}$)
5.42, 13.68, 21.93, 30.19, 38.44.
\label{fig:g337image}}

\figcaption[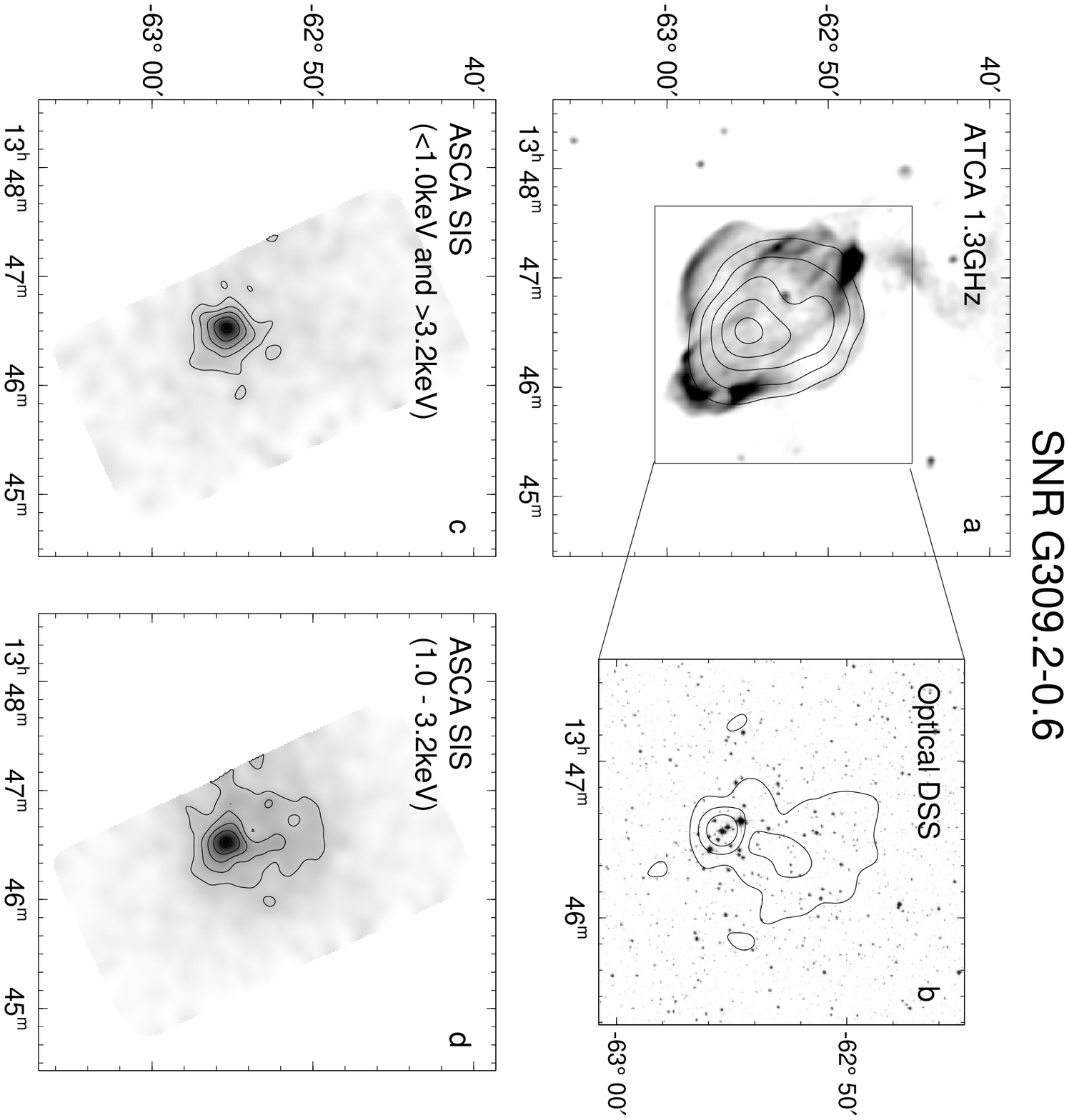]{Image of G309.2$-$0.6 in various bands. X-ray
grey-scales are linearly displayed, radio and optical grey-scales use
square root scaling. Part a: ATCA 1.3~GHz image with GIS broad-band
(0.7-10.0~keV) contours overlayed; levels (in 10$^{-4}$ counts s$^{-1}$
arcminute$^{-2}$) 2.14, 3.46, 4.77, 6.09, 7.40. Part b: Optical Digitized
Sky Survey overlayed with total {\it ROSAT} band contours: levels (in
10$^{-3}$ counts s$^{-1}$ arcminute$^{-2}$), 3.82, 6.57,
9.31. Part c: SIS 0.7-1.0~keV and 3.2-10.0~keV; contour levels (in
10$^{-3}$ counts s$^{-1}$ arcminute$^{-2}$) 0.559, 1.17, 1.79, 2.40,
3.02. Part d: SIS 1.0-3.2~keV; levels (in 10$^{-3}$ counts s$^{-1}$
arcminute$^{-2}$) 1.56, 2.78, 3.99, 5.20, 6.41.
\label{fig:g309image}}

\figcaption[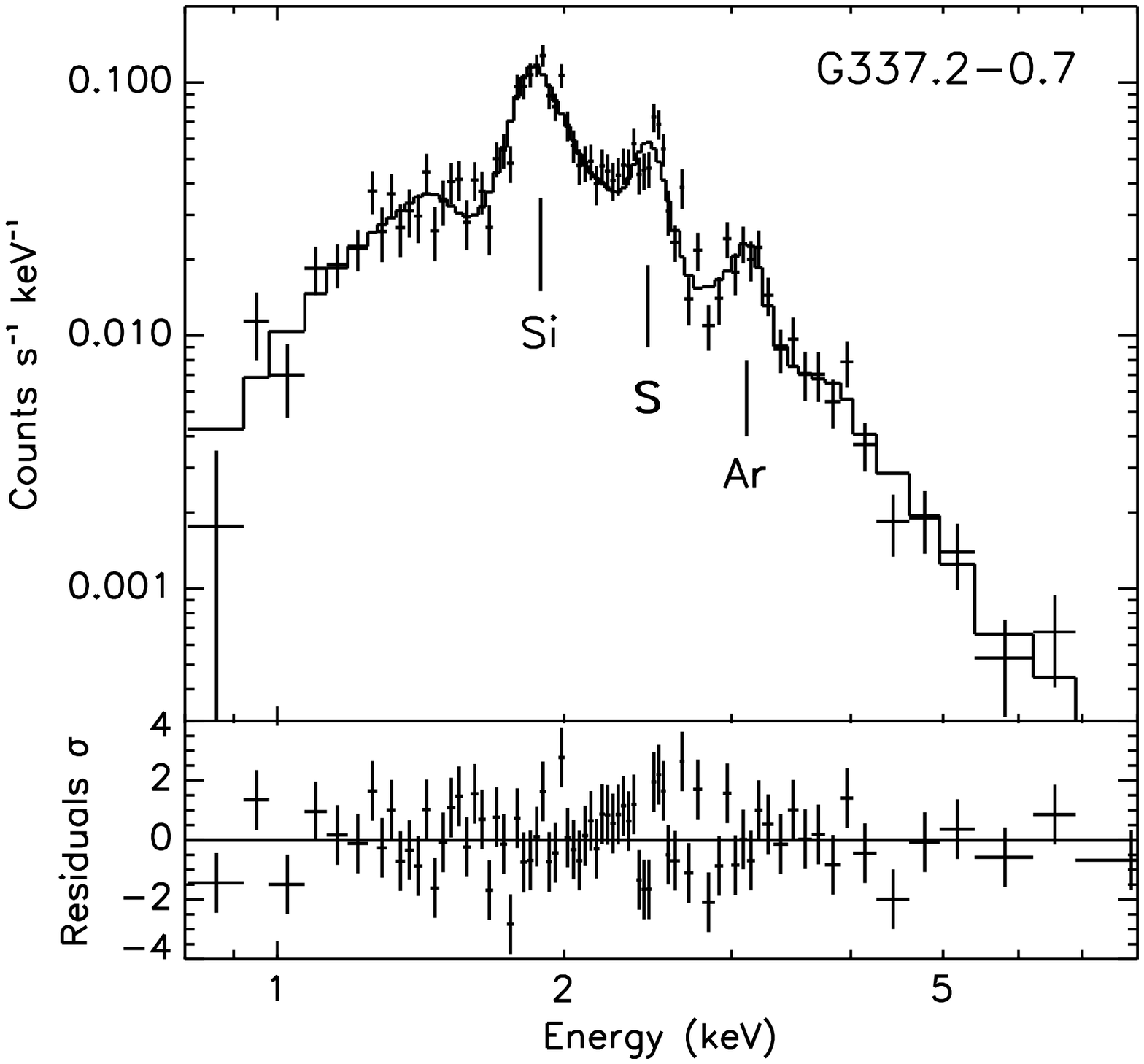]{The combined {\it ASCA} SIS0 and SIS1 spectrum
of G337.2$-$0.7 with the best-fit varying abundance model and
residuals. \label{fig:spectrum}}

\figcaption[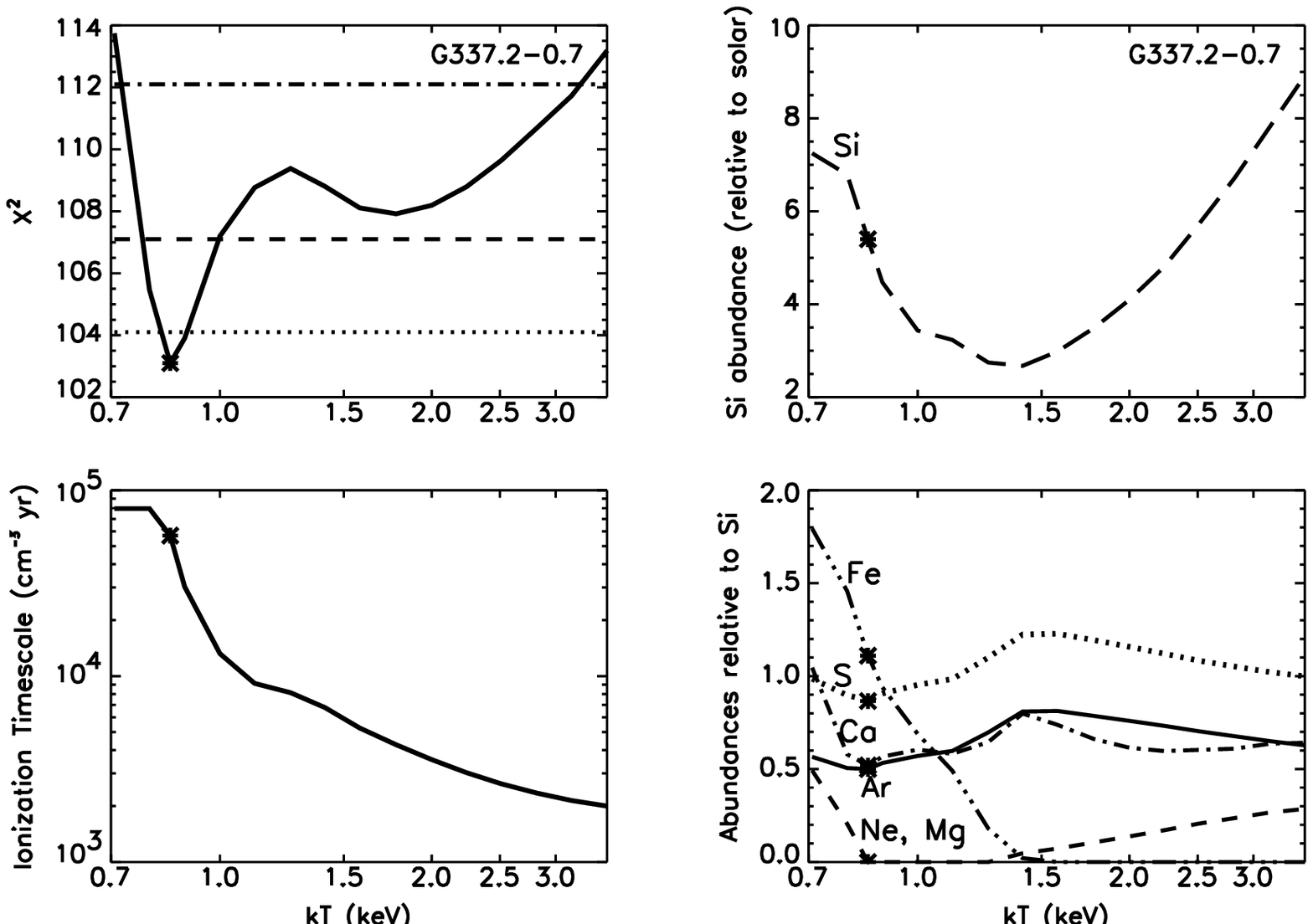]{$\chi^{2}$ and best-fit parameter values for G337.2$-$0.7
as a function of $kT$ over its 3$\sigma$ allowed range.
Relative abundances are relative to Si relative to solar.
Indicated on the $\chi^{2}$ plot are the 1, 2, and 3$\sigma$ limits
(dotted, dashes, dot-dashed lines respectively). For each parameter
the best-fit value is indicated with an asterisk.
\label{fig:kTvary}}

\figcaption[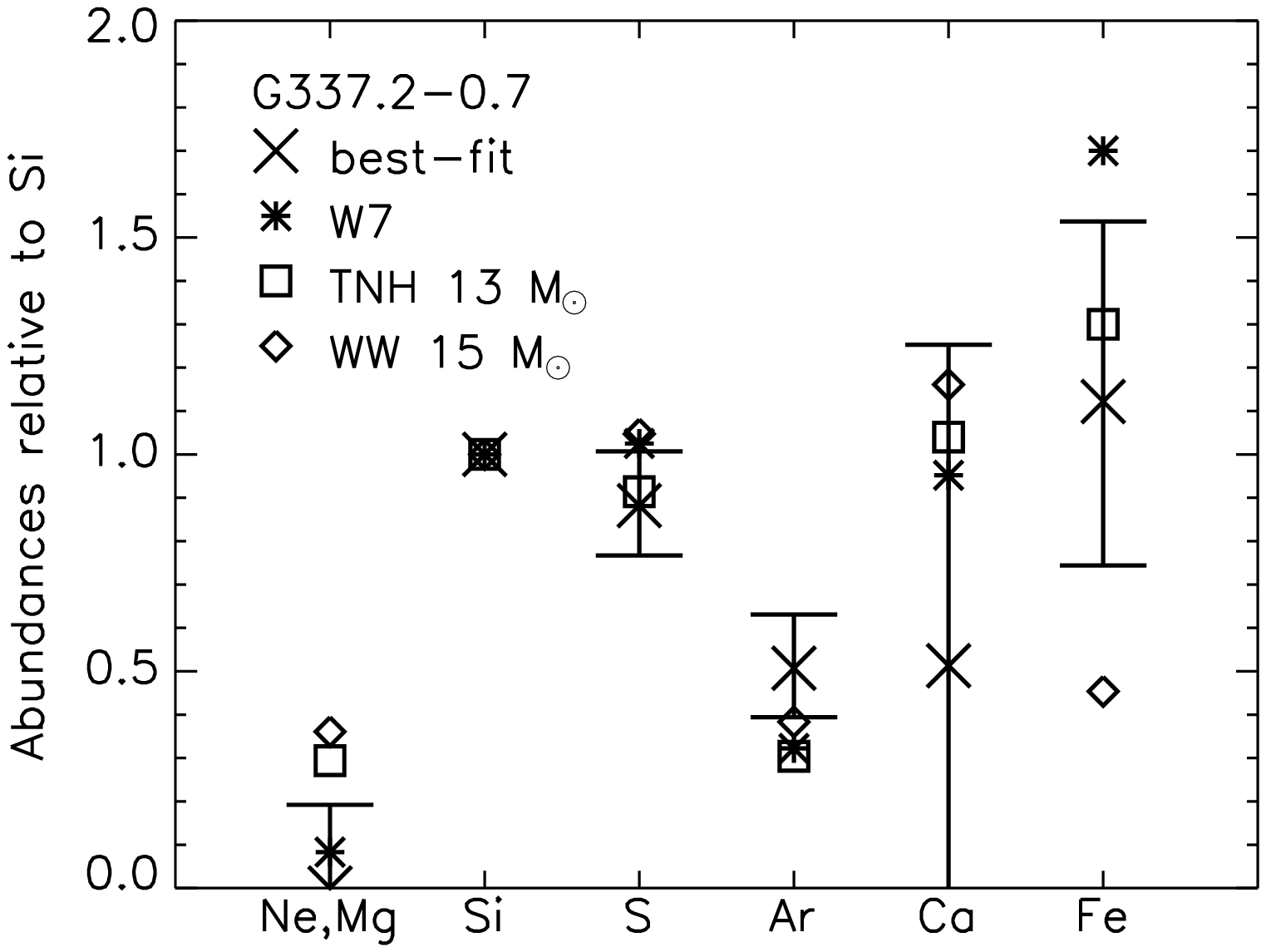]{G337.2$-$0.7: The best-fit abundances of Ne, Mg, S,
Ar, Ca, and Fe, relative to Si relative to solar, with 1$\sigma$
errors on that ratio. Also shown are the expected relative abundances
for a type Ia SNe (W7) and two type II SNe models (TNH 13~$M_{\odot}$
and WW 15~$M_{\odot}$) \label{fig:abund}}

\figcaption[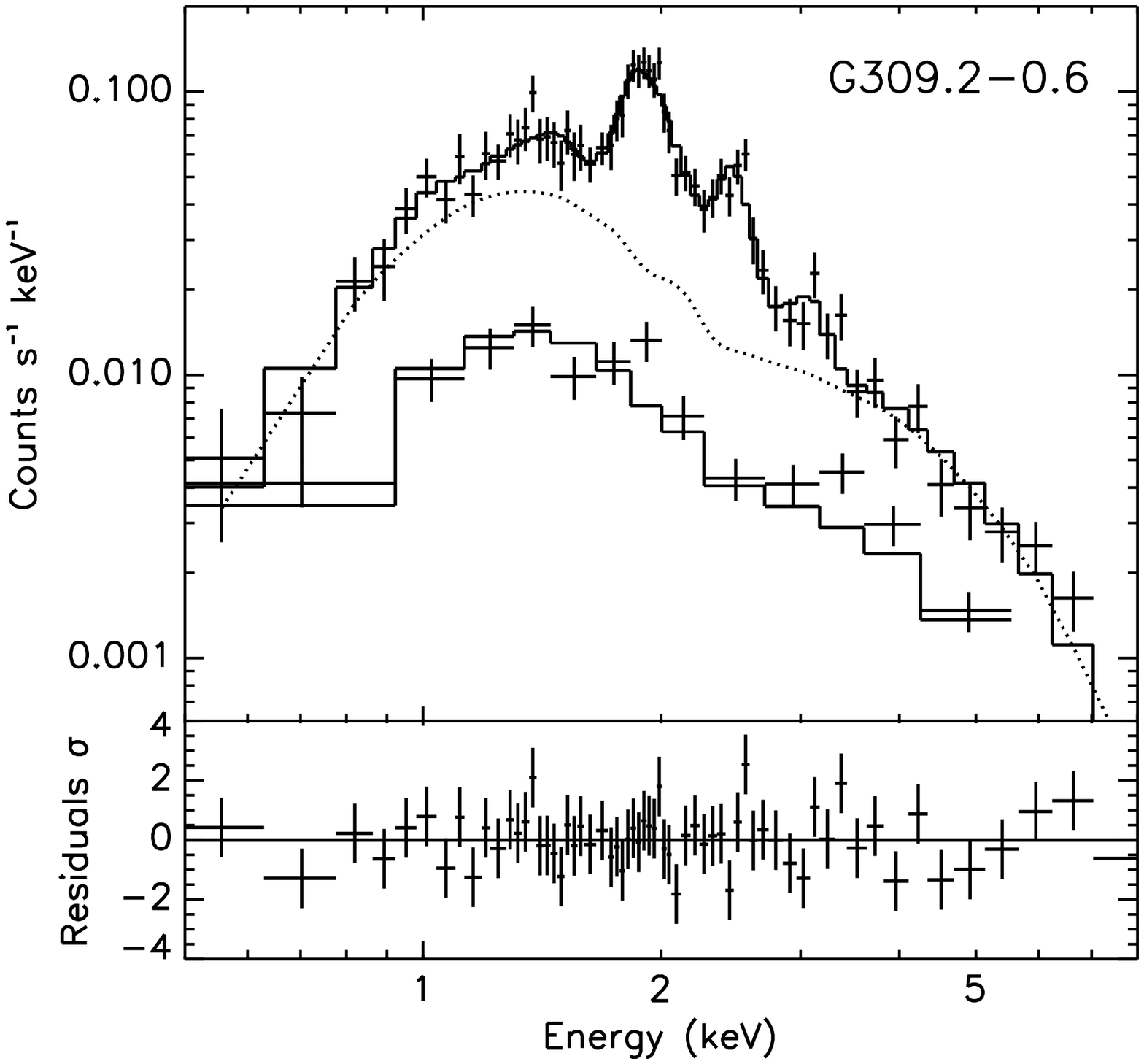]{The combined {\it ASCA} SIS0 and SIS1 spectrum of
G309.2$-$0.6 with the nominal contamination best-fit varying abundance
model and residuals. The spectrum from the hard-point source, its
model and the contamination component of the SNR model (dotted curve)
are also shown. 
\label{fig:spectrum309}}

\figcaption[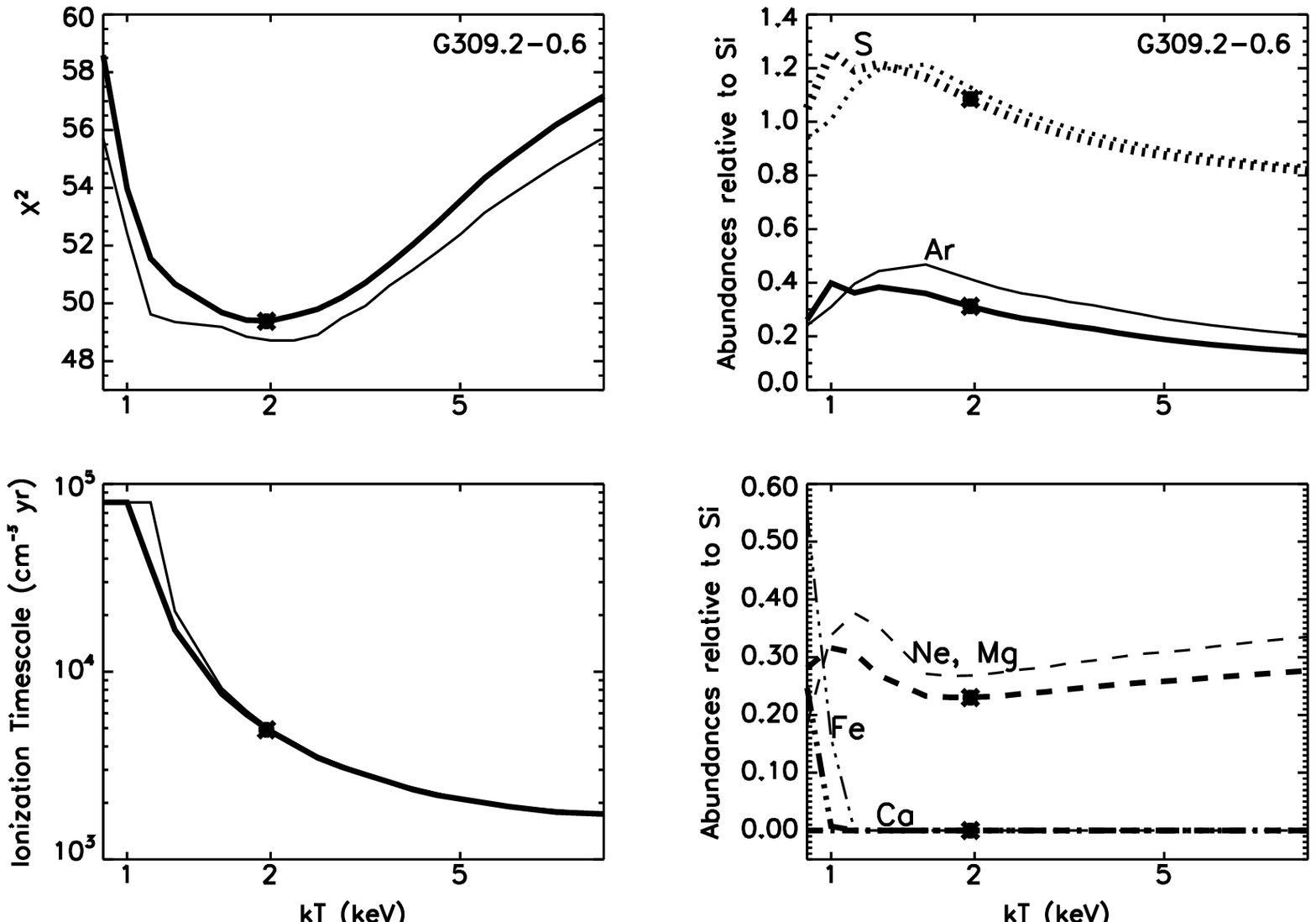]{$\chi^{2}$ and best-fit parameter values for
G309.2$-$0.6 as a function of $kT$, for both the nominal contamination
model (in bold) and the 10\% less contamination model (thin
lines). Abundances are relative to Si relative to solar. For each
parameter, the best-fit value in the nominal contamination model is
indicated with an asterisk.
\label{fig:kTvary309}}

\figcaption[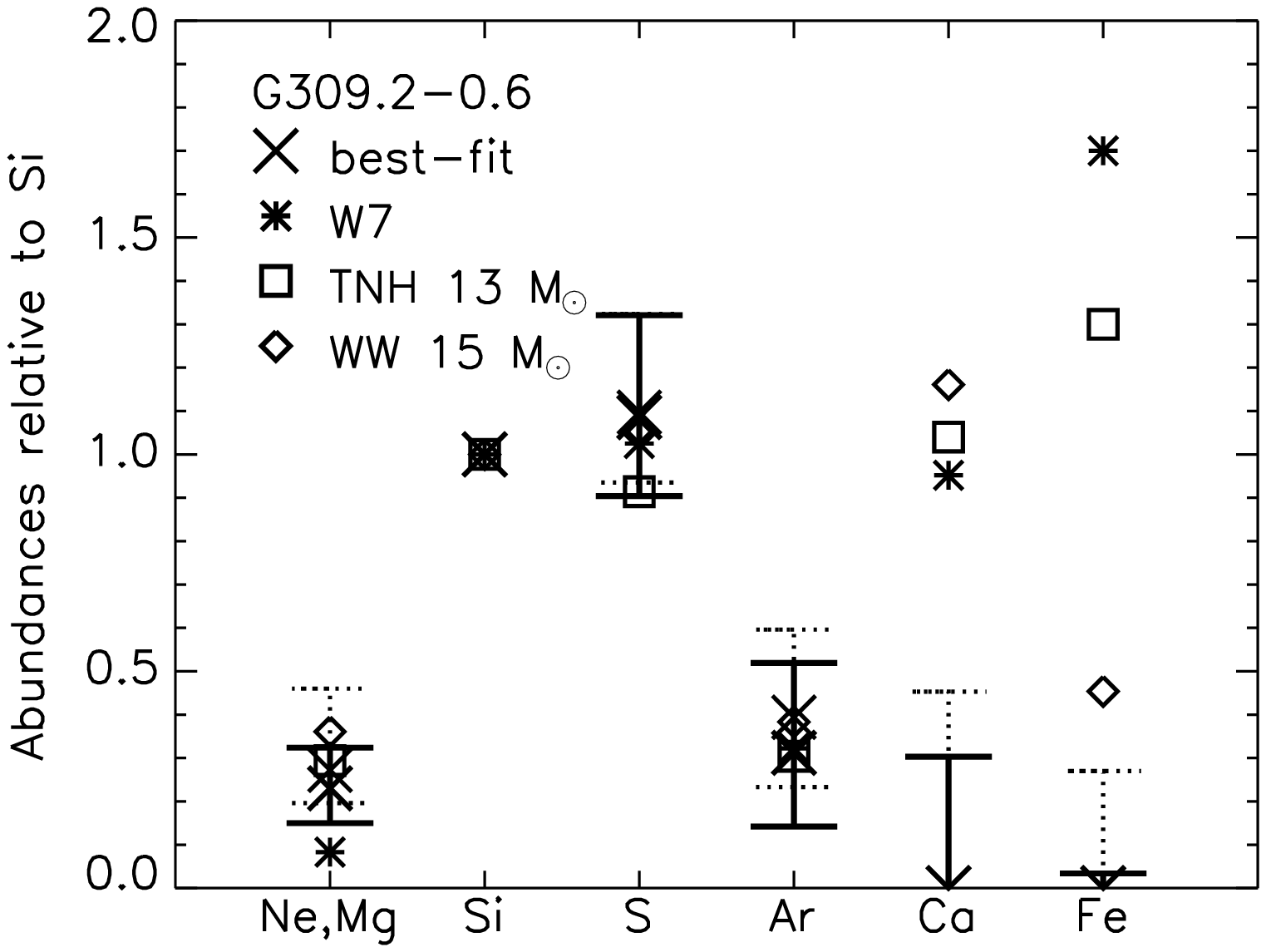]{G309.2$-$0.6: The best-fit abundances of Ne, Mg, S,
Ar, Ca, and Fe, relative to Si relative to solar, with 1$\sigma$
errors on that ratio, for both the nominal contamination model (solid)
and the 10\% less contamination model (dotted). Also shown are the
expected relative abundances for a type Ia SNe (W7) and two type II
SNe models (TNH 13~$M_{\odot}$ and WW 15~$M_{\odot}$)
\label{fig:abund309}}

\clearpage

\begin{deluxetable}{cc}
\tablewidth{0pt}
\tablecaption{Best-Fit Parameters for G337.2$-$0.7 for Varying Abundances Model\label{ta:best-fit2}}
\tablehead{
Parameter & Value
}
\startdata
$kT$ (keV)  & $0.85^{+0.04}_{-0.03}$ \\
$n_{\mathrm{e}}t$ (cm$^{-3}$ yr) & $(5.7^{+\infty}_{-3.6})\times10^{4}$ \\
$N_{\mathrm{H}}$ (atoms cm$^{-2}$)  
	& $(3.5 \pm 0.3)\times10^{22}$ \\
Emission Integral & \\
($n_{\mathrm{e}}n_{\mathrm{H}}$V)/4$\pi$D$^{2}$
	& $(1.7^{+0.9}_{-1.1})\times10^{12}$ \\
$n_{\mathrm{H}}$\tablenotemark{a} (atoms~cm$^{-3}$)
	& $1.4 \pm 0.3$ \\ 
Ne, Mg abundance\tablenotemark{b} & $0.0^{+2.1}$\\
Si abundance & $5.4^{+12.2}_{-2.2}$ \\ 
S abundance & $4.7^{+9.9}_{-1.8}$ \\
Ar abundance &$2.7^{+4.6}_{-1.1}$ \\
Ca abundance &$2.8^{+7.1}_{-2.8}$ \\
Fe abundance &$6^{+19}_{-3}$ \\ \tableline
$\chi ^{2}$ (d.o.f.)  & 103.1 (67) \\
Reduced $\chi ^{2}$ & 1.539 \\ 
\enddata
\tablenotetext{a}{Derived from the emission integral with a
nominal distance of 
10~kpc, and the volume of a thin shell of radius 3$^{\prime}$ as
seen in the radio image}
\tablenotetext{b}{Abundances relative to solar}
\end{deluxetable}

\clearpage

\begin{deluxetable}{ccc}
\tablewidth{0pt}
\tablecaption{Best-Fit Parameters for G309.2$-$0.6 for Varying Abundances Model\label{ta:best-fit4}}
\tablehead{
Parameter & \multicolumn{2}{c}{Value} \\
  & nominal contamination  & 10\% less contamination
}
\startdata
$kT$ (keV)  & $1.96^{+1.0}_{-0.6}$ & $2.11^{+0.9}_{-1.0}$ \\
$n_{\mathrm{e}}t$ (cm$^{-3}$ yr) & $(4.9^{+15}_{-2.0})\times10^{3}$ 
	& $(4.6^{+\infty}_{-1.7})\times10^{3}$ \\
$N_{\mathrm{H}}$ (atoms cm$^{-2}$)  
	& $(0.7 \pm 0.3)\times10^{22}$ & $(0.6 \pm 0.3)\times10^{22}$ \\
Emission Integral &  & \\
($n_{\mathrm{e}}n_{\mathrm{ion}}$V)/4$\pi$D$^{2}$
	& $(9.7^{+200}_{-9.7})\times10^{8}$ 
	& $(10.0^{+750}_{-10.0})\times10^{8}$ \\
$n_{\mathrm{ion}}$ \tablenotemark{a} (atoms~cm$^{-3}$)
	& $0.02^{+0.03}_{-0.01}$ & $0.02^{+0.03}_{-0.01}$ \\
Si abundance \tablenotemark{b} & $>34$     & $>11.9$  \\ 
Ne, Mg \tablenotemark{c} & $0.23^{+0.09}_{-0.08}$ & $0.27^{+0.19}_{-0.08}$\\
S \tablenotemark{c} & $1.09^{+0.24}_{-0.18}$      & $1.10^{+0.22}_{-0.17}$ \\
Ar \tablenotemark{c} & $0.31^{+0.21}_{-0.17}$      & $0.39^{+0.21}_{-0.16}$ \\
Ca \tablenotemark{c} & $0.0^{+0.30}$             &$0.0^{+0.45}$ \\
Fe \tablenotemark{c} & $0.0^{+0.03}$             &$0.0^{+0.27}$  \\ \tableline
$\chi ^{2}$ (d.o.f.)  & 49.38 (53)       & 48.59 (53) \\
Reduced $\chi ^{2}$ & 0.932              & 0.917 \\ 
\enddata
\tablenotetext{a}{Derived from the emission integral with a
nominal distance of 
5.4~kpc, and the volume of a thin shell of radius 6$^{\prime}$ as
seen in the X-ray image}
\tablenotetext{b}{Abundance relative to solar}
\tablenotetext{c}{Abundances relative to Si, relative to solar, 
in the form: (Z/Si)$\times$[(Si/H)$_{\odot}$/(Z/H)$_{\odot}$]}
\end{deluxetable}

\clearpage

\begin{figure}
\epsfig{figure=f1.ps}
\end{figure}

\clearpage

\begin{figure}
\epsfig{figure=f2.ps}
\end{figure}

\clearpage

\begin{figure}
\hspace*{1.0in}
\epsfig{figure=f3.ps}
\end{figure}

\clearpage

\begin{figure}
\epsfig{figure=f4.ps}
\end{figure}

\clearpage

\begin{figure}
\epsfig{figure=f5.ps}
\end{figure}

\clearpage

\begin{figure}
\hspace*{1.0in}
\epsfig{figure=f6.ps}
\end{figure}

\clearpage

\begin{figure}
\epsfig{figure=f7.ps}
\end{figure}

\clearpage

\begin{figure}
\epsfig{figure=f8.ps}
\end{figure}

\end{document}